\documentclass[aps, twocolumn,floats, showpacs]{revtex4}
\usepackage{amsmath,amssymb,graphicx}
\usepackage{epsfig}
\usepackage{graphicx}
\usepackage{enumerate}

\newcommand{\beq}{\begin{equation}}
\newcommand{\eeq}{\end{equation}}

\newcommand{\bea}{\begin{eqnarray}}
\newcommand{\eea}{\end{eqnarray}}

\begin{document}
\title{\footnotetext{APS copyright, to be printed in Phys. Rev. Lett.}
Single mode heat rectifier: Controlling energy flow between electronic conductors }

\author{Dvira Segal}
\affiliation{Chemical Physics Theory Group, Department of Chemistry, University of Toronto,
80 Saint George St. Toronto, Ontario, Canada M5S 3H6}

\date{\today}
\begin{abstract}
We study heat transfer between conductors, mediated by the excitation of a monomodal
harmonic oscillator. Using a simple model, we show that the onset of rectification in the system
is directly related to the nonlinearity of the electron gas dispersion relation.
When the metals have strictly linear dispersion relation a Landauer type expression for the thermal current holds,
symmetric with respect to the  temperature difference.
Rectification becomes prominent when deviations from linear dispersion exist, and the fermionic model cannot be mapped
into a harmonic- bosonized- representation.
The effects described here are relevant for understanding
radiative heat transfer and vibrational energy flow in electrically insulating molecular junctions.
\end{abstract}

 \pacs{66.70.+f, 44.10.+i, 44.40.+a, 73.23.-b}

\maketitle


Developing nanoscale devices with asymmetric conduction properties
is a long standing challenge both for fundamental science, for
resolving the microscopic mechanisms controlling transport, and for
practical applications. In the last years, rectification  in
mesoscopic and molecular level systems have attracted much
attention. Recent works demonstrated electrical rectification in
various systems, including mesoscopic semiconductor structures
\cite{E-rectifier}, one-dimensional (1D)  systems of interacting
electrons \cite{Braunecker}, and within molecular junctions
\cite{Metzger,Zhu}.
%
The growing interest in the {\it thermal} properties of nanoscale structures \cite{Rev}
turned rectification of {\it phononic} current into a topic of great interest,
with implications on nanoscale machinery \cite{thermal-machinery}
and thermal computation \cite{heat-computer}. The different setups
demonstrating this effect  \cite{Terraneo,
Casati,Rectif,Zhang,RectifE} all rely on two basic
assumptions: The device should be asymmetric with respect to the two
terminals, and the system's normal modes should nonlinearly
interact.

{\it Radiative} thermal conductance, where heat exchange between 
metals is mediated by the generation of {\it photons} is a new topic
of interest \cite{Clealand}. It was recently proved that at low
temperatures photonic heat conduction is quantized
\cite{Pekola},
setting an upper bound on single-channel information flow
\cite{Blencowe}. From the practical aspect, at the nanoscale,
radiative heat flow may compete with vibrational energy transfer,
thus these two processes must be considered for properly estimating
the thermal conductance of molecular level systems \cite{Clealand}.


In this paper we investigate monomode mediated energy exchange
between two metals, and resolve the necessary conditions for
manifesting thermal rectification.
We show that when the metallic leads have {\it linear} dispersion
relation, or in other words, when the bosonization approach can be
employed to yield a harmonic Hamiltonian, a Landauer type expression
for the energy current holds, symmetric with respect to the
temperature difference. In contrast, for conductors with nonlinear
dispersion relation,  when the approximations involving the standard
bosonization scheme break, the energy current can be rectified.
Deviations from the Tomonaga-Luttinger bosons picture \cite{TL} thus
eminently relates to the onset of rectification in the system.

\begin{figure}[htbp]
\hspace{2mm}
{\hbox{\epsfxsize=65mm \epsffile{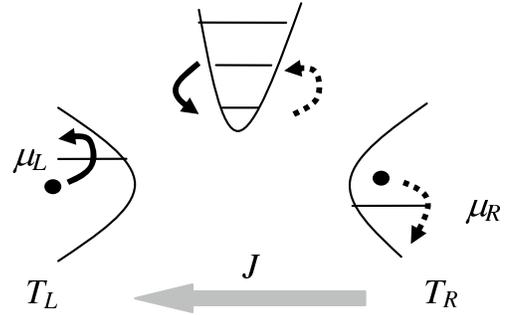}}}
\caption{A schematic representation our model.
While electron flow is blocked, even for
$\mu_L>\mu_R$, heat current is flowing right to left ($T_R>T_L$)
through an excitation of the local harmonic mode.
The curved arrows represent energy transfer processes between the 
leads and the intermediate mode.  
} \label{Fig0}
\end{figure}
%
The fermionic model consists two metallic leads held at
different temperatures, coupled by a single harmonic mode. 
The Hamiltonian includes three contributions
$H_F=H_0+H_e+V_F$,
\bea
H_0&=&\omega_0 b_0^{\dagger}b_0, \nonumber\\
H_e&=& \sum_{\nu,k} \epsilon_k c_{\nu,k}^{\dagger} c_{\nu,k},
\nonumber\\
V_F&=&\sum_{\nu,k,k'}\alpha_{\nu,k;\nu,k'} c_{\nu,k}^{\dagger}c_{\nu,k'}f.
\label{eq:Hamilton}
\eea
$H_0$ includes a single harmonic mode (subsystem) of frequency
$\omega_0$, creation operator $b_0^{\dagger}$. It can be considered
as a monomodal electromagnetic field, or in a different context, it
represents a local vibrational mode of an intermediate molecular
unit. Henceforth we refer to it as a local mode. This mode interacts
with two fermionic reservoirs ($H_e$), where $c^{\dagger}_{\nu,k}$
($c_{\nu,k}$) creates (annihilates) an electron at the $\nu=L,R$
metal with momentum $k$, disregarding the spin degree of freedom.
The oscillator-metals interaction term $V_F$ couples scattering
processes within each metal to the subsystem degrees of freedom,
where $f$ is a local mode operator. For simplicity, we use
$f=b_0^{\dagger}+b_0$, and assume that the coupling constants
$\alpha_{\nu}$ are energy independent and real numbers. Since 
 $V_F$ does not directly couple the  
reservoirs, the model (\ref{eq:Hamilton}) 
leaves out charge transfer processes,
assuming the tunneling barrier is high and the constriction is long. 
However, {\it energy} can be transferred between the two metals mediated by
the excitation  of the  local mode. For a schematic representation
see Fig. \ref{Fig0}.

While the aim of this paper is fundamental, to understand the microscopic origin of
asymmetric transport in simple fermionic Hamiltonians, the generic
model studied here may be realized in different systems: (i)
Radiative energy transfer, where the transfer of the lead's excess
energy is mediated by the excitation of electromagnetic modes. Other
mechanisms for energy transfer are neglected, e.g., thermionic
emission, vibrational energy flow, and photon tunneling. (ii)
Vibrational energy flow, where thermal energy is transferred between
two metallic terminals through a vibrating link. Here the basic
assumptions are that the molecular link electrically insulates, and
that vibrational energy cannot directly flow from the leads to the
molecule due to a large mismatch of the metal-molecule phononic
spectra.

For 1D noninteracting electrons with an unbounded {\it strictly
linear} dispersion relation, $\epsilon_k=\epsilon_F+ v_F(|k|-k_F)$,
$\epsilon_F$ is the Fermi energy and $v_F$ is the velocity at the
Fermi energy, bosonization can be employed to yield an equivalent
bosonic Hamiltonian, $H_B=H_0+H_b+V_B$. Here $H_0$ is the same as
above, while the fermionic degrees of freedom are written in terms
of Tomonaga-Luttinger (TL) bosons \cite{TL}, creation operator
$b^{\dagger}_{\nu,q}$ ($\nu=L,R$),
\bea
&& H_b=\sum_{\nu,q}\omega_qb_{\nu,q}^{\dagger}b_{\nu,q}; \,\,\,\,\ \omega_q \propto q,
 \nonumber\\
&&V_B= \sum_{\nu,q}\kappa_{\nu,q}(b_{\nu,q}^{\dagger}+b_{\nu,q}) (b_0^{\dagger}+b_0).
\label{eq:HB}
\eea
The new coupling parameters $\kappa_{\nu,q}$  relate to the
couplings  $\alpha_{\nu}$  [Eq. (\ref{eq:Hamilton})]
\cite{Chang,Leggett}. Assuming an ohmic dissipation, the spectral
function of the $\nu$ boson bath $g_{\nu}(\omega)$ can be written in
terms of the fermionic parameters,
\bea
&&  g_{\nu}(\omega)\equiv 4 \pi \sum_{q}\kappa_{\nu, q}^2\delta(\omega-\omega_q) =2 \pi \omega \xi_{\nu};
\nonumber\\
&&\xi_{\nu}=\frac{1}{2}\left[\frac{2}{\pi} {\rm atan} (\pi \rho_{\nu}(\epsilon_F) \alpha_{\nu})\right]^2,
\label{eq:BF}
\eea
with $\rho_{\nu}(\epsilon_F)$ the density of states at the Fermi energy.
The boson Hamiltonian $H_B$ is fully harmonic, and thus can be
written in terms of noninteracting collective modes.
The heat current in this model can be exactly calculated to yield a Landauer type
expression \cite{Segalcond},
\bea
J=\frac{2}{\pi}\int  {\cal T}(\omega)[n_B^L(\omega)-n_B^R(\omega)] \omega d\omega.
\label{eq:Jh}
\eea
Here $n_B^{\nu}(\omega)=[e^{\omega/T_{\nu}}-1]^{-1}$ is the
Bose-Einstein occupation factor with temperature $T_{\nu}$. The $L$
to $R$ transmission coefficient is given by ${\cal
T}(\omega)=\frac{\omega^2 \Gamma_B^L
\Gamma_B^R}{[(\omega^2-\omega_0^2)^2+(\Gamma_B^L+\Gamma_B^R)^2\omega^2]}$
\cite{Segalcond}, with the relaxation rate $\Gamma_B^{\nu}=2\pi
\sum_{q}\kappa_{\nu,q}^2\delta(\omega-\omega_q)$. In the weak
coupling limit, $\Gamma_B^{\nu}<\omega_0$, the transmission
coefficient is sharply peaked around $\omega_0$, and Eq.
(\ref{eq:Jh}) reduces into a resonant energy transfer expression,
\bea
J= \omega_0\frac{\Gamma_B^L\Gamma_B^R}{\Gamma_B^L+\Gamma_B^R}[n_B^L(\omega_0)-n_B^R(\omega_0)].
\label{eq:Jb}
\eea
Here $\Gamma_B^{\nu}$ is calculated at the (local oscillator)
frequency $\omega_0$. For weak coupling, $\pi
\rho(\epsilon_F)\alpha<1$, Eq. (\ref{eq:BF}) yields
$\xi=2\rho(\epsilon_F)^2\alpha^2$, and the relaxation rate is given explicitly by
\bea
\Gamma_B^{\nu}=2\pi \omega_0 \alpha_\nu^2 \rho^2_{\nu}(\epsilon_F).
\label{eq:GammaB}
\eea
Eqs. (\ref{eq:Jh})-(\ref{eq:GammaB}) describe the heat current of
the model Hamiltonian (\ref{eq:Hamilton}) under the assumption that
the two fermionic reservoirs have a strictly linear dispersion
relation. It is clear that in this case thermal rectification cannot
show up, since exchanging the temperatures of the leads simply
reverses the sign of the heat current (\ref{eq:Jh}), not the
absolute value \cite{comment}.
We show next that when the bosonization method
cannot be trivially employed, i.e. when the dispersion relation is
not linear, the model (\ref{eq:Hamilton}) can bring in an
interesting rectifying behavior.

Assuming a general dispersion relation, the dynamics of the model (\ref{eq:Hamilton}) is
analyzed in the weak coupling system-bath limit.
Going into the Markovian limit, the probabilities $P_n$ to occupy the $|n\rangle$ state
of the local oscillator satisfy the master equation
\bea
\dot P_n= \sum_m P_m k_{m \rightarrow n}-P_n\sum_m k_{n\rightarrow m},
\label{eq:popul}
\eea
where the transition rate from the local oscillator state
$|m\rangle$ to $|n\rangle$ is additive in the $L$ and $R$
reservoirs, $k_{n\rightarrow m}=k^L_{n\rightarrow
m}+k^R_{n\rightarrow m}$, due to the linear form of the interaction
[Eq. (\ref{eq:Hamilton})] \cite{Rectif}. In steady state, the heat
current across the system is given by (calculated e.g. at the $L$
side),
\bea
J=\sum_{m,n}E_{m,n}P_nk^L_{n\rightarrow m},
\label{eq:current}
\eea
with $E_{m,n}=E_m-E_n$. 
At the level of the Golden Rule formula, the transition rates are given by
\bea
&&k_{n\rightarrow m}^\nu =
2\pi|f_{m,n}|^2
 \sum_{k,k'} |\alpha_{\nu,k;\nu,k'}|^2
\nonumber\\
&&\times
n_F^{\nu}(\epsilon_k) [1-n_F^{\nu}(\epsilon_{k'})] \delta(\epsilon_{k}-\epsilon_{k'}-E_{m,n})
\nonumber\\
&&=2\pi | f_{m,n}|^2 \int d \epsilon n_F^{\nu}(\epsilon) [1-n_F^{\nu}(\epsilon-E_{m,n})]
F_{\nu}(\epsilon).
\nonumber\\
&&=-2 \pi |f_{m,n}|^2 n_B^{\nu}(E_{m,n})
\nonumber\\
&&\times
\int d\epsilon
\left[ n_F^{\nu}(\epsilon)-n_F^{\nu}(\epsilon-E_{m,n}) \right]F_{\nu}(\epsilon).
\label{eq:FGR}
\eea
Focusing on the last equality, interestingly we see that the thermal
properties of the reservoirs are concealed both within  the
Fermi-Dirac distribution function
$n_F^{\nu}(\epsilon)=[e^{(\epsilon-\mu_{\nu})/T_{\nu}}+1]^{-1}$ and
the Bose-Einstein occupation factor
$n_B^{\nu}(\epsilon)=[e^{\epsilon/T_{\nu}}-1]^{-1}$. It is therefore
clear that when the integral yields a temperature independent constant, the
statistic of the reservoirs is fully bosonic. The other elements in
(\ref{eq:FGR}) are the matrix elements of the system operator
$f_{m,n}=\langle m|f|n\rangle$, and the dimensionless interaction
term
%
$F_{\nu}(\epsilon)=|\alpha_{\nu}|^2\rho_{\nu}(\epsilon)\rho_{\nu}(\epsilon-E_{m,n})$,
%
which encloses the properties of the reservoirs multiplied by the
system-bath couplings $\alpha_{\nu}$. These
couplings might be taken different at the two ends, bringing in a spatial
asymmetry.

We assume next that the density of states slowly varies in the energy window $E_{m,n}$.
The interaction function is then expanded around the chemical potential \cite{Persson},
\bea
F_{\nu}(\epsilon)\approx F_{\nu}(\mu_{\nu})+ \gamma_{\nu} \frac{|\epsilon|-\mu_{\nu}}{\mu_{\nu}},
\label{eq:F}
\eea
with $\gamma_{\nu}$ a dimensionless number of order unity. Using
this form, the integration in Eq. (\ref{eq:FGR}) can be 
performed when the Fermi energies are much bigger than the
conduction band edge, $\mu_{\nu} \gg E_c$ \cite{Persson}. Making use
of the following relationships,
\bea
&&\int_{-\infty}^{\infty} d \epsilon [n_F^{\nu}(\epsilon)-n_F^{\nu}(\epsilon-E_{m,n})] \approx -E_{m,n},
\nonumber\\
&&\int_{-\infty}^{\infty} |\epsilon| d \epsilon [n_F^{\nu}(\epsilon)-n_F^{\nu}(\epsilon-E_{m,n})]
\nonumber \\
&&\approx -1.4 E_{m,n} T_{\nu}; \,\,\,\,\, (T_{\nu}>|E_{m,n}|),
\eea
we get
\bea k^{\nu}_{n \rightarrow m}= 2 \pi |f_{m,n}|^2  n_{B}^{\nu}(E_{m,n})E_{m,n}
\left [F_{\nu}(\mu_{\nu}) +  1.4 \gamma_{\nu} \frac{T_{\nu}}{\mu_{\nu} }
\right].
\nonumber\\
\label{eq:kPersson}
\eea
Note that $n_{B}(-E_{m,n})=-[n_B(E_{m,n})+1]$, thus the excitation
and relaxation rates induced by the $\nu$ reservoir satisfy
the detailed balance relation, $k_{n\rightarrow
m}^{\nu}/k_{m\rightarrow n}^{\nu} = e^{-E_{m,n}/T_{\nu}}$.
We consider next the limit of a constant density of states, taking
$\gamma=0$. Eq. (\ref{eq:kPersson}) then becomes
\bea
&& k_{n\rightarrow n-1}^{\nu} =n \Gamma_F^{\nu}(\omega_0)
[1+n_B^{\nu}(\omega_0)];
\nonumber\\
&&\Gamma_F^{\nu}(\omega_0)=2 \pi F_{\nu}(\mu_{\nu})\omega_0,
\label{eq:rate}
\eea
where the bilinear interaction form was employed
$f=(b_0^{\dagger}+b_0)$, leading to nearest neighbors transitions
only. We can also calculate the states population in steady state by
putting $\dot P_n=0$ in Eq. (\ref{eq:popul}) \cite{Rectif},
\bea
P_n=x^n(1-x); \,\,\,\,
x=\frac{ \sum_{\nu}\Gamma_F^{\nu}(\omega_0)n_{B}^{\nu}(\omega_0)}{ \sum_{\nu}\Gamma_{F}^{\nu}(\omega_0)[1+n_B^{\nu}(\omega_0)] }.
\label{eq:popul2}
\eea
Finally, the heat current for the $\gamma=0$ case is calculated with
the help of Eq. (\ref{eq:current}),
\bea J=\omega_0 \frac{\Gamma_F^{L}
\Gamma_F^{R}}{\Gamma_F^L+\Gamma_F^R} [n_B^L(\omega_0) -
n_B^R(\omega_0)].
\label{eq:harmonic}
\eea
$\Gamma_F^{\nu}$ is calculated at the frequency
$\omega_0$. In the linear dispersion limit we thus recover the
resonant energy behavior  (\ref{eq:Jb}) obtained with the equivalent boson
Hamiltonian (\ref{eq:HB}). Note that the rates calculated with the
different methods, $\Gamma_F^{\nu}$ and $\Gamma_B^{\nu}$, are equal,
see Eqs. (\ref{eq:GammaB}) and (\ref{eq:rate}). 

We evaluate next the current when $\gamma\neq 0$, i.e. for a model
with an energy dependent density of states. 
For the linear coupling case, $f=b_0^{\dagger}+b_0$, 
only transitions between nearest levels
survive and Eq. (\ref{eq:kPersson}) becomes
%
$k_{n \rightarrow n-1}^{\nu}=n \Gamma_F ^{\nu}(\omega_0) [1+n_B^{\nu}(\omega_0)]
\left( 1+ \lambda_{\nu} \frac{T_{\nu}} {\mu_{\nu}} \right)$,
%
where $\lambda_{\nu}=\frac{1.4 \gamma_{\nu}} {F_{\nu}(\mu_{\nu})}$.
The steady state population is therefore
given by Eq. (\ref{eq:popul2}) with $\Gamma_F^{\nu} \rightarrow \Gamma_F^{\nu}(1+\lambda_{\nu}T_{\nu}/\mu_{\nu})$.
Next we calculate the energy current using Eq. (\ref{eq:current}).
In the classical limit ($T_{\nu}>\omega_0$) we obtain
\bea
J=\frac{\Gamma_F^L\Gamma_F^R(1+\lambda_L\frac{T_L}{\mu_L})(1+\lambda_R\frac{T_R}{\mu_R})}{\Gamma_F^L(1+\lambda_L\frac{T_L}{\mu_L})+\Gamma_F^R(1+\lambda_R\frac{T_R}{\mu_R})}
(T_L-T_R).
\label{eq:Jan}
\eea
The dimensionless parameter $\lambda_{\nu}$ effectively quantifies the deviation from the linear dispersion
case. System-bath interactions $\alpha_{\nu}$ are contained within the coefficients $\Gamma_F^{\nu}$
computed at the subsystem frequency $\omega_0$.

Comparing Eq. (\ref{eq:Jan}) to the high temperature limit of Eq.
(\ref{eq:harmonic}) interestingly reflects the deviations from
the TL boson picture in transport properties. Eq.
(\ref{eq:Jan}) has three important characteristics: First, the heat
current obtained is a nonlinear function of the temperature
difference. In fact, for $\Gamma_F^L\gg \Gamma_F^R$,
$J\propto a_1\Delta T +\lambda a_2 \Delta T ^2 $. The constants 
$a_{1,2}$ depend on subsystem and bath parameters, $\Delta T=T_L-T_R$. Second,
this expression manifests nontrivial controlability over
the energy current by tuning the chemical potentials.
Thirdly, this result demonstrates thermal
rectification, as the thermal current is different when switching
the temperature bias, assuming some asymmetry is included, e.g. the
chemical potentials, or the local mode-bath interactions are
different at the two terminals. When the electronic properties of
the reservoirs are equivalent, $\lambda \equiv\lambda_{\nu}$,
$\mu_a\equiv\mu_{\nu}$, we find 
%
$\Delta J=J_{+\Delta T}+J_{-\Delta T}
\propto  -\lambda \frac{ \Delta T^2 }{\mu_a}(\Gamma_F^L-\Gamma_F^R)$,
where  $J_{\pm \Delta T}=J(T_L-T_R=\pm \Delta T)$. The
proportionality factor is given by $G\approx
\Gamma_F^L\Gamma_F^R/(\Gamma_F^L+\Gamma_F^R)^{2}$ for $\lambda
T_{\nu}/\mu_a<1$. Thus, a system made of a local {\it harmonic} mode
coupled {\it asymmetrically} to two metals with {\it energy
dependent density of states}, rectifies heat. Rectification becomes
more effective with increasing $\lambda$, i.e. when the density of
states strongly varies with energy.
Therefore, the same parameter that "quantifies" the departure of the fermion model from the boson picrture, determines
the effectiveness of rectification in the system.
%
\begin{figure}[htbp]
{\hbox{\epsfxsize=70mm \epsffile{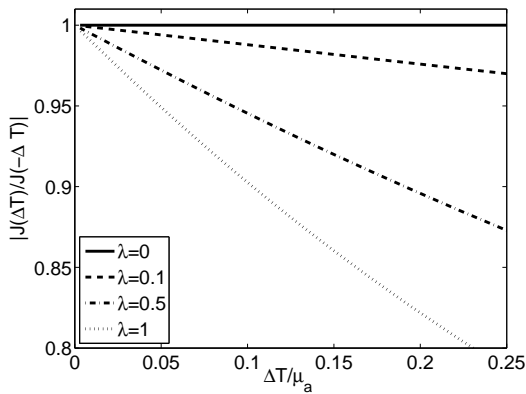}}}
\caption{Thermal rectification with increasing  deviations
from constant density of states.
$\lambda=0$ (full); $\lambda=0.1$ (dashed); $\lambda=0.5$ (dashed-dotted);
$\lambda=1$ (dotted).
$\Gamma_F^{L}=0.5$, $\Gamma_F^{R}=0.02$, $\mu_a=\mu_{\nu}=1$,
temperature of the cold bath  is $T_{low}$=0.1.
}
\label{Fig1}
\end{figure}
\begin{figure}[htbp]
{\hbox{\epsfxsize=70mm \epsffile{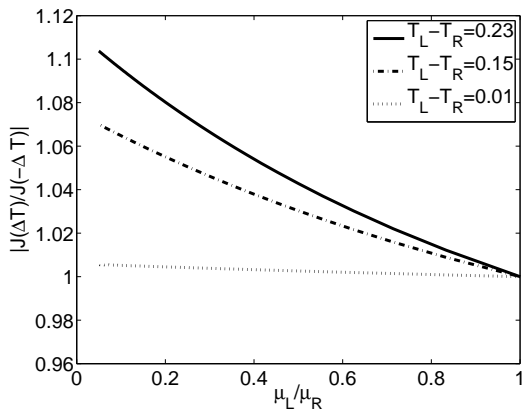}}}
\caption{Thermal rectification due to chemical potential difference.
$\Delta T=0.23$ (full); $\Delta T=0.15$ (dashed); $\Delta T=0.01$ (dotted).
$\Gamma_F^L=\Gamma_F^R=0.1$,
$\mu_L=1$, $\lambda_L=\lambda_R$=1.
}
\label{Fig2}
\end{figure}
%
%
Figure \ref{Fig1} presents this effect. We plot the absolute value of the ratio $J_{+\Delta T}/J_{-\Delta T}$,
and find that it linearly departs from unity with increasing
temperature difference and $\lambda$, in agreement with the analytical expression ($T_a=\frac{T_L+T_R}{2}$),
\bea
|J_{+\Delta T}/J_{-\Delta T}| \approx
1- \frac{\lambda}{\mu_a} \Delta T
\frac{\Gamma_F^L-\Gamma_F^R}{\left(\Gamma_F^L+\Gamma_F^R\right)(1+\frac{\lambda T_a}{\mu_a})}.
\label{eq:Jpm}
\eea
Figure \ref{Fig2} demonstrates the dependence of the rectifying
behavior on the leads' chemical potentials for a system 
symmetrically coupled to the terminals using Eq. (\ref{eq:Jan}).
We find that at small temperature difference rectification is negligible, while for
$\Delta T/\mu_{L}>0.1$ the asymmetry in the current can be large up
to $10\%$. We note however that tunneling of electrons becomes a
significant source of noise at large voltage bias and/or temperature
differences.
Interestingly, the energy current can be also controlled by tuning $\Delta\mu$, 
an outcome of the energy dependence of the density of states.
%
%

The thermal current can be calculated for other models besides (\ref{eq:F}),
expressing  Eq. (\ref{eq:FGR}) as
$k_{n\rightarrow m} ^{\nu}=-2 \pi |f_{m,n}|^2 n_B^{\nu}(E_{m,n}) \alpha_{\nu}^2 g_{\nu}(T_{\nu})$,
where $g_{\nu}(T_{\nu})$ is defined through this relation.
Eq. (\ref{eq:Jpm}) then reduces to
${\mathcal R}\equiv |J_{+\Delta T}/J_{-\Delta T}|\approx g_R(T_R)/g_R(T_L)$ for the linear coupling model
if $\alpha_{L} \gg \alpha_R$. 
We consider a superconducting $R$ lead ($L$ can be a normal metal), as it 
effectively suppresses normal electronic thermal conductance  \cite{Rev,Pekola},
and numerically evaluate $g_R(T)$ with the density of states
$\rho_R(\epsilon)=\frac{|\epsilon|}{\sqrt{\epsilon^2-\Delta^2}}\Theta(|\epsilon|-\Delta)$.
Using $\Delta=0.2$ meV (aluminum), $\omega_0=0.4$ meV, and $T=0.08-0.2$ meV, we get
$g_R(T)\sim a_1 +a_2T$ where $a_2 T \gg a_1$.
The rectification ratio then becomes ${\mathcal R}\sim T_R/T_L$, 
which can be as large as 2.5 for the above parameters,
potentially measurable using the setup of Ref. \cite{Pekola2}.
Similarly, a narrow bandgap (0.1 eV) semiconducting lead yields ${\mathcal R}\sim 10$ for
$\omega_0=0.1$ eV at the temperature range $T\sim 0.01-0.5$ eV.

Previous studies on {\it phononic} heat transfer in 1D chains
demonstrated that anharmonic interactions are crucial for
manifesting thermal rectification
\cite{Terraneo,Casati,Rectif,Zhang}. Our result complies with
this observation: A fermionic reservoir with a nonlinear dispersion
relation can be represented by a bath of bosons comprising of
nonlinear interactions \cite{Haldane}. The fermionic
Hamiltonian (\ref{eq:Hamilton}) can thus be mapped into a fully
bosonic model describing a harmonic link bilinearly connected to
{\it anharmonic} thermal baths. 

In summary, we present here a simple model of single-mode  energy
transfer between metals. The model can describe 
energy flow through vibrating link and radiative heat transfer.
We resolve the microscopic requirements for manifesting rectification:
For noninteracting electrons, 
rectification appears when the reservoirs' 
density of states,  or analogously,  system-bath couplings,
are energy dependent. 
The effects discussed here are not limited to the specific form used 
(\ref{eq:F}), and are a general manifestation of the breakdown of the TL boson picture.
Since dissipative reservoirs typically contain anharmonic interactions, finite rectification
of the energy current between metals and dielectric surfaces is an inevitable effect.

\begin{acknowledgments}
I thank J. G. Groshaus for helpful comments and the
University of Toronto for financial support.
\end{acknowledgments}



\begin{thebibliography}{32}
\bibitem{comment}
An anharmonic mode asymmetrically coupled to linearly dispersed metals
can rectify heat.
For a two-level (anharmonic) subsystem,
$J=\omega_0\Gamma_B^L \Gamma_B^R[\Gamma_B^L(1+2n_B^L) +\Gamma_B^R(1+2n_B^R)]^{-1} (n_B^L-n_B^R)$ \cite{Rectif}, see
Eq. (\ref{eq:Jb}) for definitions.
\end{thebibliography}
\end{document}